\documentclass[conference]{IEEEtran}
\IEEEoverridecommandlockouts

\usepackage{cite}
\usepackage{amsmath,amssymb,amsfonts}
\usepackage{graphicx}
\usepackage{textcomp}
\usepackage{xcolor}
\def\BibTeX{{\rm B\kern-.05em{\sc i\kern-.025em b}\kern-.08em
    T\kern-.1667em\lower.7ex\hbox{E}\kern-.125emX}}

\usepackage{tcolorbox}
\usepackage[utf8]{inputenc} 
\usepackage[T1]{fontenc}    
\usepackage{hyperref}       
\usepackage{enumitem}
\usepackage{url}            
\usepackage{booktabs}       
\usepackage{amsfonts}       
\usepackage{nicefrac}       
\usepackage{microtype}      
\usepackage{xcolor}         
\usepackage{graphicx}
\usepackage{multicol}
\usepackage{multirow}
\usepackage{color, colortbl}
\usepackage{amsbsy}
\usepackage{array, makecell} %
\usepackage{boldline} 
\usepackage{hhline}
\usepackage[labelfont=bf]{caption}
\usepackage{todonotes}
\usepackage{algorithm}
\usepackage{algpseudocode}
\newcolumntype{P}[1]{>{\centering\arraybackslash}p{#1}}
\newcolumntype{M}[1]{>{\centering\arraybackslash}m{#1}}
\newcolumntype{N}{@{}m{0pt}@{}}

\definecolor{LightGray}{gray}{0.9}
\definecolor{mygrey}{rgb}{0.6,0.6,0.6}

\begin{document}

\title{SEAL: Speaker Error Correction using Acoustic-conditioned Large Language Models}

\author{\IEEEauthorblockN{Anurag Kumar\textsuperscript{\dag*}}
\IEEEauthorblockA{\textit{Ohio State University}\\
kumar.1109@osu.edu}
\and
\IEEEauthorblockN{Rohit Paturi\textsuperscript{*}}
\IEEEauthorblockA{\textit{AWS AI Labs}\\
paturi@amazon.com}
\and
\IEEEauthorblockN{Amber Afshan}
\IEEEauthorblockA{\textit{AWS AI Labs}\\
afsamber@amazon.com}
\and
\IEEEauthorblockN{Sundararajan Srinivasan}
\IEEEauthorblockA{\textit{AWS AI Labs}\\
sundarsr@amazon.com}
\thanks{\textsuperscript{\dag}Work performed during internship at AWS AI Labs.}
\thanks{\textsuperscript{*}These authors contributed equally to this work.}
}

\maketitle
\begin{abstract}
Speaker Diarization (SD) is a crucial component of modern end-to-end ASR pipelines. Traditional SD systems, which are typically audio-based and operate independently of ASR, often introduce speaker errors, particularly during speaker transitions and overlapping speech. Recently, language models including fine-tuned large language models (LLMs) have shown to be effective as a second-pass speaker error corrector by leveraging lexical context in the transcribed output. In this work, we introduce a novel acoustic conditioning approach to provide more fine-grained information from the acoustic diarizer to the LLM. We also show that a simpler constrained decoding strategy reduces LLM hallucinations, while avoiding complicated post-processing. Our approach significantly reduces the speaker error rates by 24-43\% across Fisher, Callhome, and RT03-CTS datasets, compared to the first-pass Acoustic SD.

\end{abstract}

\section{Introduction}
Speaker Diarization (SD) solves the problem of determining “Who spoke when” in an audio recording. SD systems can be broadly categorized into two types: 1) Modular systems \cite{wan2018generalized, dawalatabad2021ecapa}, which consist of components like a segmenter and an embedding model , and 2) End-to-end systems \cite{fujita2019end, kinoshita2021integrating, li24x_interspeech}, which are designed to handle speech overlaps and are directly optimized for diarization by incorporating permutation invariant training loss, followed by a clustering phase . 
However, many real-world applications like meeting analytics, call-center analytics, and video captioning often require associating spoken words with the speaker labels, as opposed to just the speaker time ranges predicted by SD modules.

Conventional ASR \cite{wang2020transformer,yang2022conformer, rekesh2023fast, radford2023robust} systems are generally designed for speaker-agnostic scenarios and answer the question “What was spoken” without providing speaker labels. Multi-speaker transcription systems, however, address the question “Who spoke what and when,” which is essential for many practical applications. Various approaches have been developed to achieve this, primarily falling into three categories: 1) Speech/Speaker Separation followed by ASR \cite{9413423,paturi22_interspeech,zhao2023mossformer}, 2) Speaker-attributed ASR \cite{kanda22_interspeech,shi23d_interspeech,sklyar2021streaming, berger23_interspeech}, and 3) Modular ASR and SD systems. Challenges with Speech Separation and SA-ASR systems include difficulties with a larger number of speakers \cite{9413423}, duplicated artifacts \cite{paturi22_interspeech}, lack of speaker timestamps \cite{kanda22_interspeech}, and handling long-form audio \cite{paturi22_interspeech}. Independent ASR and SD systems, which operate separately and later reconcile their results, offer a more practical solution for multi-speaker transcription pipelines.

In modular ASR-SD systems, the ASR and SD components are trained independently and combined using their respective timestamps, a process we refer to as reconciliation. This reconciliation can lead to speaker attribution errors, compounded by inherent errors in the SD system, especially during speaker transitions or in overlapping speech regions \cite{paturi23_interspeech, Paturi2024}. Although End-to-End Neural Diarization (EEND) systems can handle speech overlaps, since the speaker agnostic ASR system can only detect words from one speaker at a time, this can still produce speaker errors in overlap regions.

Lexical information, which complements acoustic data, can be crucial for accurately assigning speaker labels \cite{paturi23_interspeech, Paturi2024, park2018multimodal, xia2022turn}. For example, by analyzing a transcript like "how are you i am good", one can infer a speaker change between "how are you" and "i am good". In this work, we propose \textbf{SEAL}, a \textbf{S}peaker \textbf{E}rror Corrector using \textbf{A}coustic-conditioned \textbf{L}arge Language Models, which integrates the LLM's inherent lexical knowledge with acoustic information from the SD system. Additionally, we introduce a Constrained Decoding (CD) technique to prevent the LLM from hallucinating, ensuring that the generated word sequence strictly matches the original transcript without modifying any words.

\section{Related Work}

Some earlier works \cite{shafey2019joint, adedeji2024sound} have utilized lexical knowledge for speaker diarization, but they have been limited to speakers with specific roles. More recently, a Lexical Speaker Error Correction (SEC) framework \cite{paturi23_interspeech} and its extension with acoustic score fusion \cite{Paturi2024} have been proposed to correct general speaker errors using a pre-trained Language Model (LM) encoder, such as RoBERTa, demonstrating good improvements. However, these methods only utilize a pre-trained LM encoder and do not take advantage of the capabilities of modern instruction-tuned large LLMs.

A few works \cite{park2023enhancing, wang2024diarizationlmspeakerdiarizationpostprocessing, efstathiadis2024llm} have explored the use of LLMs for either performing SD or correcting speaker errors. For instance, \cite{park2023enhancing} uses an LLM to predict speaker probabilities for the next word, incorporating these into beam search decoding along with the acoustic scores for speaker diarization. However, this approach will be computationally expensive since the LLM must be invoked for every word in the transcript, and the beam search adds additional complexity. Other works \cite{wang2024diarizationlmspeakerdiarizationpostprocessing, efstathiadis2024llm} employ LLMs to correct speaker diarization errors from an acoustic SD module by prompting a fine-tuned LLM to rectify errors in the transcript. A limitation of this approach is that the LLM lacks access to acoustic information in the audio which can lead to over or under-correction as highlighted in \cite{Paturi2024}. 
Additionally, a challenge with LLM-based speaker correction is word hallucination. To address this, \cite{wang2024diarizationlmspeakerdiarizationpostprocessing} introduces the "Transcript Preserving Speaker Transfer" (TPST) algorithm, which maps predicted speaker labels back to the input word sequence. While this removes any word hallucinations in the final output, it doesn't eliminate the LLM's inherent tendency to hallucinate, potentially leading to sub-optimal speaker assignments.

\textbf{SEAL} builds on the strengths of \cite{Paturi2024} and \cite{wang2024diarizationlmspeakerdiarizationpostprocessing}, proposing a method to condition the LLM with acoustic speaker scores for enhanced error correction. Inspired by previous works \cite{geng2023grammar, willard2023efficientguidedgenerationlarge, das2024speechverse} that apply various constraints to guide LLM generation, we implement a similar Constrained Decoding (CD) strategy, which prevents LLM word hallucinations by ensuring that generated word sequences strictly match the original transcripts. Our results demonstrate that these approaches lead to significant improvements in speaker error correction.

\begin{figure}[t]
  \centering
  \includegraphics[width=\columnwidth]{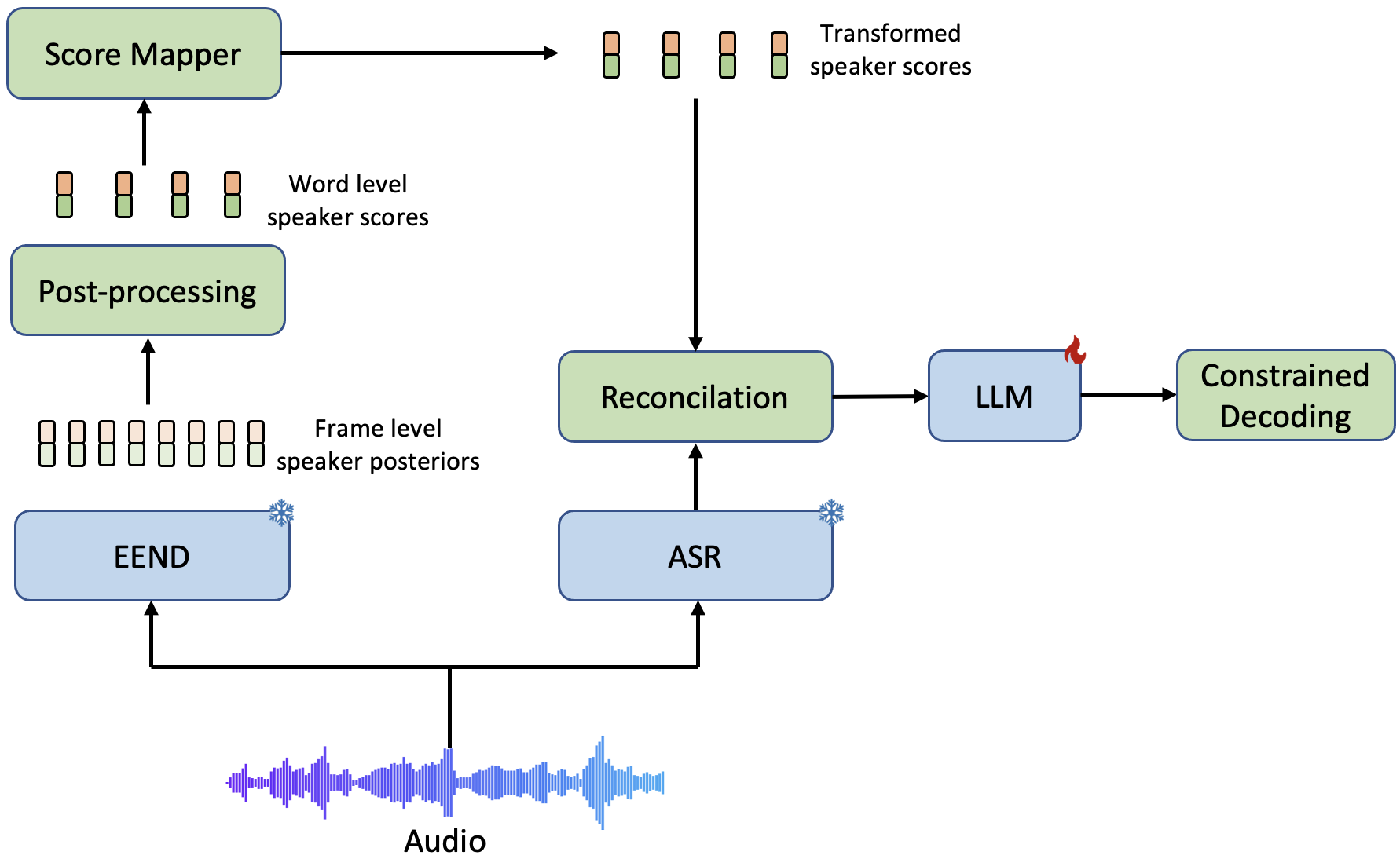}
  \caption{Proposed framework of SEAL.}
  \vspace{-1\baselineskip}
  \label{fig1}
\end{figure}

\section{\textbf{S}peaker \textbf{E}rror Correction using \textbf{A}coustic-conditioned \textbf{L}LMs (\textbf{SEAL})}
An overview of the \textbf{SEAL} framework is shown in Fig 1. 
\subsection{First pass Speaker Diarization Module}

EEND is a preferred choice as a first pass acoustic diarization module for our setup as it can efficiently handle overlaps and output soft speaker scores that can acoustically ground the LLM decisions for speaker error correction. Given the frame level acoustic features $X = \{x_i\}_{i=1}^{t}$, $X \in R^{t \times f}$, where $t$ and $f$ represent the number of frames and feature dimensions respectively, the speaker posteriors $P = \{p_i\}_{i=1}^{t}$, $P \in R^{t \times k}$, where k represents the number of speakers, are obtained using an EEND network $f_{EEND}$

\begin{equation}
    \{ p_1, p_2, ..., p_t\} = f_{EEND}(x_1, x_2, ..., x_t)
\end{equation}

These frame-level posterior scores are then median-filtered and mean-pooled at the word level to produce word-level aggregated posteriors $\{s_i\}_{i=1}^{W}$, where $W$ is the number of words in the transcript, with values ranging from [0, 1], following the approach in \cite{Paturi2024}.


\begin{figure}[!t]
 \centering
 \includegraphics[width=0.8\linewidth]{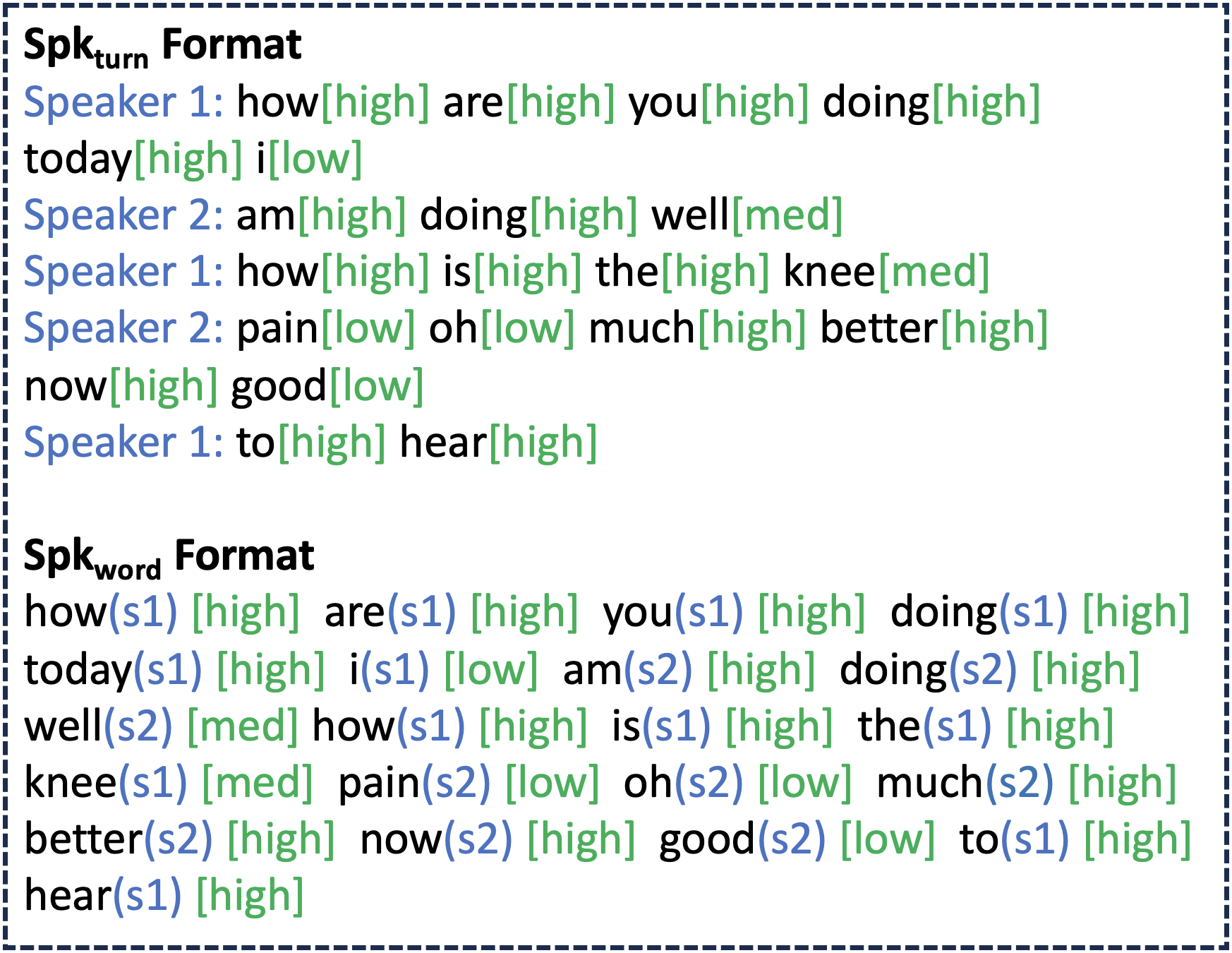}
 \caption{Different input transcript formats with acoustic score mapped to 3 labels: low, med, high.}
 \vspace{-1\baselineskip}
 \label{fig2}
\end{figure}
\subsection{Acoustic Conditioning}
\label{Score Integration}
To account for the fact that LLMs are predominantly trained on text and often struggle with understanding numerical values without specialized mechanisms \cite{imani2023mathprompter}, we propose converting the soft scores from the EEND module into more easily interpretable labels. We classify the scores into three categories: “low”, “medium” and “high”. For each word-level score $s_i$, we assign a category 
$c_i$ based on the following formula: 
\[
  c_i = \begin{cases}
    high, & \text{for } Th_{med} < s_i \leq 1 \\
    med, & \text{for } Th_{low} < s_i \leq Th_{med} \\
    low, & \text{for } 0 < s_i \leq Th_{low}
  \end{cases}
\]
The thresholds $Th_{low}$, $Th_{med}$ are tuned based on a dev set and we refer this component as the “Score Mapper” module in Figure 1. To assess the effectiveness of this approach, we also conducted an ablation study using the raw probabilities from the EEND module as well as a more granular categorization approach that divides the scores into 10 uniformly distributed integer classes between [0,1]. We refer to the model finteuned only on ASR transcripts without any acoustic conditioning as LLM\textsubscript{AC-none}, the model acoustically conditioned with raw speaker probabilities $s_i$ as LLM\textsubscript{AC-prob}, the model acoustically conditioned with speaker confidence labels $c_i$ as LLM\textsubscript{AC-label} and the LLM conditioned on the 10 integer classes as LLM\textsubscript{AC-int}. These comparisons are detailed in Section \ref{results_section}.

With these different conditioning strategies, we experimented with two transcript formats: “Spk\textsubscript{turn}”, which follows the same format from \cite{wang2024diarizationlmspeakerdiarizationpostprocessing, efstathiadis2024llm} where speaker turns are indicated after the speaker label, and “Spk\textsubscript{word}” where each word is followed by its speaker label and corresponding acoustic label. These formats are illustrated in Fig \ref{fig2}.

\subsection{Constrained Decoding (CD)}

Fine-tuned LLMs may still hallucinate and produce non-verbatim outputs \cite{wang2024diarizationlmspeakerdiarizationpostprocessing, efstathiadis2024llm}. To address this, \cite{wang2024diarizationlmspeakerdiarizationpostprocessing} introduces the "Transcript Preserving Speaker Transfer" (TPST) algorithm, which maps predicted speaker labels to the input word sequence. While this approach prevents word hallucinations in the final output, it does not stop the LLM from generating hallucinations during inference. This limitation affects speaker assignments since TPST doesn't leverage lexical context from the input or generated transcripts.

To maintain output fidelity while correcting speaker errors, we propose a Constrained Decoding (CD) strategy that limits the LLM's predictions to specific words at each step. The permitted words are either the next word or a speaker label in the Spk\textsubscript{turn} format, and only speaker labels in the Spk\textsubscript{word} format, as the LLM only needs to predict speaker labels preceding each word. CD guarantees that the LLM adheres to the input word sequence while adjusting speaker labels, preventing word hallucinations. Algorithm \ref{alg:CD} provides details of this strategy.

\begin{algorithm}
\caption{Constrained Decoding Algorithm}
\label{alg:CD}
  \textbf{Inputs} \\
  \hspace*{\algorithmicindent} Input sequence of length $M:$ $w_{inp} = \{w\_in_i\}_{i=0}^{M}$ \\
  \hspace*{\algorithmicindent} LLM model$:$ $LLM()$ \\
  \hspace*{\algorithmicindent} Allowed words module$:$ $get\_allowed\_words()$ \\
  \textbf{Outputs} \\
  \hspace*{\algorithmicindent} Output sequence of length $O:$ $w_{out} = \{w\_o_i\}_{i=0}^{O}$ \\
    \begin{algorithmic}[1]
    \Procedure{ConstrainedDecoding}{}
    \State $w_{out} \gets ``"$ \textcolor{mygrey}{\Comment{{Initialize $w_{out}$ to empty string}}}
    \For{$1 \leq j \leq N$} \textcolor{mygrey}{\Comment{{$N$ is max supported seq length}}}
        \State $curr\_inp\_seq \gets \{w\_in_{i}\}_{i=0}^{j}$
        \State $next\_out\_dist \gets LLM(curr\_inp\_seq)$
        \State $allowed\_words \gets get\_allowed\_words(i, w_{inp})$
        \For{each $word$ in $next\_out\_dist$}
            \If{$word$ not in $allowed\_words$}
                \State $next\_out\_dist[word] \gets \text{-inf}$ 
            \EndIf
        \State $w\_o_j \gets \text{argmax}(next\_out\_dist)$
        \If {$w\_o_j$ is \text{<EOS>}}
            \State break
        \EndIf
        \State $w_{out} \gets w_{out} + w\_o_j$
        \EndFor
    \EndFor
    \EndProcedure
    \end{algorithmic}
\end{algorithm}

\section{Experimental Setup}
\subsection{Dataset and Metrics}
We use the Fisher dataset \cite{40LDC, 41LDC} to finetune LLMs. The same pre-processing as \cite{Paturi2024, wang2024diarizationlmspeakerdiarizationpostprocessing} was done to get a single channel Fisher audio inputs and respective ground-truth transcripts. We held out the same partition of the dataset as test set as mentioned in \cite{wang2022highly, Paturi2024, wang2024diarizationlmspeakerdiarizationpostprocessing}. In addition to that, we also test our models on CALLHOME American English (CHAE) \cite{39LDC} and RT03-CTS \cite{42LDC}. 

\begin{table}[t]
\centering
\caption{Ablation of acoustic conditioning variants for Spk\textsubscript{turn} format with TPST. Model names are defined in Section \ref{Score Integration}} 
\label{tab2}
\begin{tabular}{@{}lcccc@{}}

\toprule
  & \multicolumn{4}{c}{\textbf{CHAE}} \\ \cmidrule(l){2-5} 
  \textbf{Models} &
  \multicolumn{2}{c}{\textbf{Dev}} &
  \multicolumn{2}{c}{\textbf{Test}}\\ \cmidrule(l){2-5} 
 &
  \textbf{cpWER} &
  \textbf{$\Delta$cp} &
  \textbf{cpWER} &
  \textbf{$\Delta$cp}
   \\ \toprule

\textbf{LLM\textsubscript{AC-none}} & 17.6 & 7.47 & 14.88  & 5.89 \\
\textbf{LLM\textsubscript{AC-prob}} & 18.2 & 8.07 & 15.15 & 6.16 \\
\textbf{LLM\textsubscript{AC-int}} & 17.4 & 7.27 & 14.72 & 5.73 \\
\textbf{LLM\textsubscript{AC-label}} & \textbf{15.17} & \textbf{5.09} & \textbf{13.75} & \textbf{4.83} \\ 
\bottomrule  
\end{tabular}
\vspace{-1\baselineskip}
\end{table}

We use the concatenated minimum permutation WER or cpWER \cite{watanabe2020chime} and $\Delta$cp \cite{park2023enhancing} as our main evaluation metrics which appropriately captures the combined ASR and Speaker Diarization related errors.
\begin{table*}[htp]
\centering
\caption{Evaluation of SEAL models with other baselines. LLM\textsubscript{AC-none} is the LLM without acoustic-conditioning, LLM\textsubscript{AC-label} is the SEAL model with speaker scores mapped to labels. Best scores are bolded and second best are underlined.}
\label{tab1}
\begin{tabular}{@{}lcccccccccccc@{}}

\toprule
  & \multirow{3}{*}{\parbox{1.25cm}{\centering \textbf{Transcript\\Format}}} & \multirow{3}{*}{\parbox{1.95cm}{\centering \textbf{Decoding/\\Post-processing}}} & \multicolumn{2}{c}{\textbf{Fisher}}  & \multicolumn{4}{c}{\textbf{CHAE}} & \multicolumn{4}{c}{\textbf{RT03-CTS}} \\ \cmidrule(l){4-13} 
  \textbf{Models} & & &
  \multicolumn{2}{c}{\textbf{Test}} &
  \multicolumn{2}{c}{\textbf{Dev}} &
  \multicolumn{2}{c}{\textbf{Test}} &
  \multicolumn{2}{c}{\textbf{Dev}} &
  \multicolumn{2}{c}{\textbf{Test}} \\ \cmidrule(l){4-13} 
 &  & &
  \textbf{cpWER} &
  \textbf{$\Delta$cp} &
  \textbf{cpWER} &
  \textbf{$\Delta$cp} &
  \textbf{cpWER} &
  \textbf{$\Delta$cp} &
  \textbf{cpWER} &
  \textbf{$\Delta$cp} &
  \textbf{cpWER} &
  \textbf{$\Delta$cp} \\ \toprule
\textbf{Acoustic SD} & - & - & 12.53 & 3.72 & 15.52 & 5.44 & 14.33  & 5.41  & 9.96    & 3.57   & 10.03   & 3.87  \\
\textbf{LSEC}\cite{paturi23_interspeech} & - & - & 11.62 & 2.81 & 15.15 & 5.07 & 13.51  & 4.59  & 9.17 & 2.78 & 9.15 & 2.99  \\
\textbf{LLM\textsubscript{AC-none}} & \textbf{Spk\textsubscript{turn}} & \textbf{TPST} & 11.2  & 2.71 & 17.5 & 7.37 & 14.7 & 5.71 & 11.1 & 4.06 & 10.83 & 4.66   \\ \hline
\textbf{AG-LSEC}\cite{Paturi2024} & - & - & \underline{10.95} & \underline{2.14} & 14.15 & 4.07 & \textbf{12.95} & \textbf{4.03} & \underline{8.7} & \underline{2.31} & \textbf{8.36}  & \textbf{2.2}  \\
\textbf{LLM\textsubscript{AC-label}} & \textbf{Spk\textsubscript{turn}} & \textbf{TPST} & 11.73 & 2.88 & 15.17 & 5.09 & 13.47 & 4.37 & 9.77 & 3.38 & 9.25 & 3.09 \\
\textbf{LLM\textsubscript{AC-label}} & \textbf{Spk\textsubscript{turn}} & \textbf{CD} & 11.29 & 2.49 & \underline{14.08} & \underline{4.01} & \underline{13.1} & \underline{4.18}  & 9.35 & 2.95 & 8.75 & 2.59  \\ 
\textbf{LLM\textsubscript{AC-label}} & \textbf{Spk\textsubscript{word}} & \textbf{CD} & \textbf{10.92} & \textbf{2.12} & \textbf{13.6} & \textbf{3.53} & 13.6 & 4.68  & \textbf{8.45} & \textbf{2.05} & \underline{8.5} & \underline{2.34}  \\ \bottomrule \hline 
\end{tabular}
\vspace{-0.5\baselineskip}
\end{table*}

\begin{figure*}[htp]
  \centering
  \includegraphics[width=0.7\linewidth]{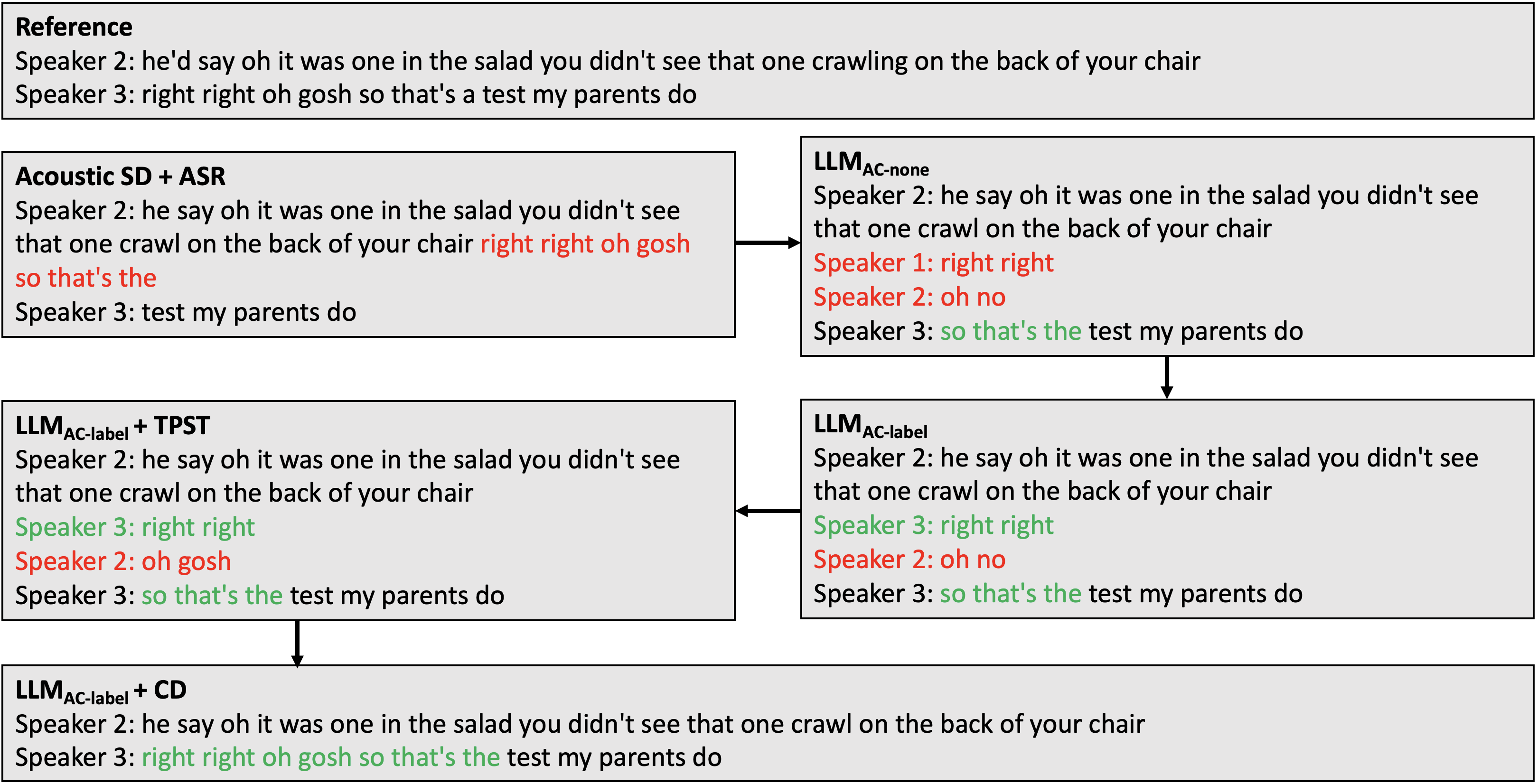}
  \caption{A Qualitative example of the incremental speaker error corrections with each of the proposed strategies. Errors are shown in red and the incremental corrections in green.}
  \vspace{-0.9\baselineskip}
  \label{fig3}
\end{figure*}
\subsection{Baseline}
We use the same baseline EEND and ASR system as reported in \cite{Paturi2024}. The EEND model consists of 6 stacked self-attention-based Transformer layers, 8 attention heads with a hidden size of 256 and 1024 internal units in the position-wise feed-forward layer. The ASR system comprises of a Conformer Acoustic model \cite{gulati2020conformer} and a n-gram Language model. 

We chose the publicly available Mistral 7b Instruct v0.2 \cite{jiang2023mistral} as the backbone LLM for this work similar to \cite{efstathiadis2024llm}. We fine-tune the LLM without any confidence scores and refer to it as the LLM\textsubscript{AC-none} which serves as our finetuning baseline. For decoding, we use the TPST algorithm from \cite{wang2024diarizationlmspeakerdiarizationpostprocessing} as our decoding baseline.

\subsection{Training}
The LLM fine-tuning was done using 4 A10 GPUs each with a VRAM of 24GB. We set our batch size to 4, gradient accumulation steps to 5 and a maximum sequence length to 1000. We used Quantized Low-Ranked Adaption of Language Models \cite{dettmers2024qlora} for efficient fine-tuning with a rank of 64 similar to \cite{efstathiadis2024llm}, to balance computational resources with model performance improvements. Similar to \cite{wang2024diarizationlmspeakerdiarizationpostprocessing, Paturi2024}, the input transcripts to the LLM are transcribed and diarized by the ASR, SD systems and the outputs that the LLMs are optimized on, are the Reference Speaker transferred version of the input transcripts (referred as Oracle transcripts in \cite{wang2024diarizationlmspeakerdiarizationpostprocessing}). This is to ensure that the LLMs produce verbatim transcripts and only correct the speaker labels.

During fine-tuning, each batch contained chunks limited to 64 words. The chunks were normalized, converted to lowercase, and stripped of punctuation before being processed by the LLM. The thresholds for acoustic score integration, $Th_{low}$ and $Th_{med}$, were set at 0.5 and 0.8, respectively, following optimization on the Fisher dev set.




\section{Results and Discussion}
\label{results_section}

We evaluate different acoustic-conditioning variants of the \textbf{SEAL} models in Table \ref{tab2}. As expected, it can be observed that LLM\textsubscript{AC-prob} struggles with raw probabilities, leading to degraded $\Delta$cp compared to the non acoustic-conditioned model, LLM\textsubscript{AC-none}. Although LLM\textsubscript{AC-int} shows marginal improvements, LLM\textsubscript{AC-label} yields the best results by converting speaker probabilities into simple, parsable words.

Based on these findings, Table \ref{tab1} benchmarks the best-performing \textbf{SEAL} model, LLM\textsubscript{AC-label}, against other baselines. The first part of the table presents models without acoustic conditioning and the first-pass Acoustic SD baseline. All systems share the same SD and ASR components, resulting in identical WERs, as word hallucinations with the LLMs are eliminated using TPST or CD. The LLM\textsubscript{AC-none} with Spk\textsubscript{turn} format replicates the DiarizationLM framework \cite{wang2024diarizationlmspeakerdiarizationpostprocessing} but with a different LLM (Mistral 7b Instruct v0.2) and baseline ASR, SD systems. While LLM\textsubscript{AC-none} improves error correction on the Fisher test set, it underperforms on other datasets, likely due to differing conversation types (informal vs. formal) and limited domain adaptability. LSEC \cite{paturi22_interspeech} performs best among text-only models, underscoring the limitations of LLMs in handling natural conversations without further conditioning.

The second part of Table \ref{tab1} compares acoustic-conditioned models using CD or TPST for generating verbatim transcripts. AG-LSEC \cite{Paturi2024} consistently performs well across datasets, leading in two out of the five data splits. Acoustic conditioning significantly improves LLM performance across all datasets, as demonstrated by the LLM\textsubscript{AC-label} rows. CD further enhances performance significantly compared to TPST as demonstrated by the TPST and CD rows with Spk\textsubscript{turn} transcript format. Additionally, changing the transcript format from Spk\textsubscript{turn} to Spk\textsubscript{word} boosts performance in most data splits. Overall, non acoustic-conditioned LLMs lag behind specialized LSEC \cite{paturi23_interspeech} and AG-LSEC \cite{Paturi2024} models, while \textbf{SEAL} models outperform AG-LSEC in most splits, achieving relative $\Delta$cp improvements of 24-43\% over first-pass Acoustic SD. Qualitative examples of the different correction strategies are shown in Figure \ref{fig3}.

\section{Conclusion}
We introduce \textbf{SEAL}, a novel \textbf{S}peaker \textbf{E}rror Corrector using \textbf{A}coustic-conditioned \textbf{L}LMs that incorporates acoustic information from the SD system, coupled with a constrained decoding strategy to ensure that generated word sequences align precisely with the original transcripts. Our evaluations across three datasets shows that acoustic conditioning significantly enhances the LLM’s Speaker Error Correction (SEC) capabilities and its ability to generalize to diverse conversation styles. The combination of our acoustic-conditioning and constrained decoding consistently surpasses both the first-pass acoustic SD baseline and the non acoustic-conditioned LLMs with TPST on the Fisher, Callhome, and RT03-CTS datasets. While \textbf{SEAL} models also achieve modest improvements over the specialized LSEC models, they offer the additional advantage of being versatile for tasks beyond SEC and will continue to improve as more powerful LLMs are developed. Future work will focus on expanding this framework to languages beyond English.

\bibliographystyle{IEEEtran}
\bibliography{IEEEabrv,refs}

\end{document}